\documentclass[prd,showpacs,showkeys,preprintnumbers,amsmath,amssymb,floatfix]{revtex4-2}
\usepackage[utf8]{inputenc}
\usepackage{hyperref}
\usepackage[dvipsnames]{xcolor}
\hypersetup{
    colorlinks=true,
    linkcolor=MidnightBlue,  %
    citecolor=RoyalBlue,     %
    urlcolor=Cerulean        %
}
\usepackage{dcolumn}%
\usepackage{bm}%
\usepackage{graphicx}
\usepackage{subfigure}
\usepackage{diagbox}

\def\be{\begin{equation}}
\def\ee{\end{equation}}
\def\bestar{\begin{equation*}}
\def\eestar{\end{equation*}}

\DeclareMathOperator{\arccoth}{arccoth}

\DeclareMathAlphabet{\mathbbold}{U}{bbold}{m}{n}

\newcommand{\Tr}{\mathrm{Tr}}

\begin{document}

\title{Phase transitions of a cubic Casimir matrix model on the fuzzy sphere}

\author{Edgar Guzmán-González}
\email{edgar.guzman@hainanu.edu.cn}
\affiliation{School of Physics and Optoelectronic
Engineering, Hainan University, 570228 Haikou, P. R. China}

\author{Marco Maceda}
\email{mmac@xanum.uam}
\affiliation{Departamento de F\'{i}sica \\ Universidad Aut\'{o}noma Metropolitana - Iztapalapa\\
                  Av. San Rafael Atlixco 186, C.P. 09340, México City, M\'exico}

\author{Andrés Tornero}
\email{andrestornero1@gmail.com}
\affiliation{Departamento de F\'{i}sica \\ Universidad Aut\'{o}noma Metropolitana - Iztapalapa\\
                   Av. San Rafael Atlixco 186, C.P. 09340, México City, M\'exico}

\date{\today}

\begin{abstract}
In this work, we analyse the phase transitions of a higher derivative matrix model on the fuzzy space, where the kinetic term is a cubic polynomial on the Casimir invariant. The phase transitions of the model are determined numerically and the presence of a triple point is exhibited. The associated critical values of the coupling parameters are calculated numerically and compared with the linear and quadratic Casimir models, showing that the cubic Casimir kinetic term helps to reduce the UV/IR mixing in these models.

\keywords{matrix model, phase transition, higher-order kinetic term}
\end{abstract}

\pacs{}

\maketitle

\section{Introduction}
\label{intro}

Matrix models play an important role in understanding field theory at the
classical and quantum levels. For example, they provide a formulation in terms
of finite-dimensional matrix degrees of freedom while preserving the symmetries
relevant to the physical system under consideration. The physical behaviour of
these models may then be explored through their phase structure and the
associated phase transitions. If the symmetry properties of the model are
sufficiently simple, analytic expressions for several observables, such as the
free energy, can often be obtained. Notably, if the action depends
exclusively on the eigenvalues, the model can be reduced to an integral over the
eigenvalue degrees of freedom. In this case, the potential terms are treated
exactly, allowing for substantial analytic progress. However, once a kinetic
term such as a Laplacian in the matrix sense is included, this reduction can
no longer be performed in closed form. A common way to overcome this difficulty
is to treat the kinetic contribution perturbatively, leading to effective
multitrace models~\cite{Steinacker:2005wj,OConnor:2007ibg,Samann:2014hjr,Ydri:2021cam}. In this framework, the
effects of the kinetic term are encoded in an effective action containing
products of traces of matrix powers. The partition function may then be analysed
using large-$N$ saddle-point methods or character expansions based on trace
identities for $SU(N)$ matrices~\cite{OConnor:2007ibg}. These approaches provide
controlled approximations, although their validity depends on the structure of
the underlying kinetic operator and may become limited in more general settings.

Of particular interest are matrix models defined on fuzzy spaces for several
reasons. Firstly, they can provide insight into the regularisation of
singularities in classical field theories, as illustrated by the
Grosse–Wulkenhaar model for $\phi^4$ theory~\cite{Grosse:2004yu,Grosse:2005da}. Furthermore, such
models have found applications in high-energy physics, quantum gravity, and
emergent geometry. Examples include the six-matrix model introduced
in~\cite{Ydri:2016osu,Ydri:2016kua}, which provides a mechanism for describing
different phases of the universe, as well as studies of phase transitions in the
Grosse–Wulkenhaar model~\cite{Prekrat:2023fts}.

Techniques from random matrix theory~\cite{Livan:2018} have also proven useful
in this context, and many analytical methods are adapted from that framework.
However, the intrinsic complexity of these backgrounds often limits the
applicability of standard approaches. As discussed above, perturbative
treatments are frequently required, and in many cases even these become
insufficient.

As a result, semi-analytical and numerical methods are essential for exploring
the phase structure of these models~\cite{GarciaFlores:2009hf}. This is
particularly relevant when studying observables that require significant
computational resources, such as tunnelling probabilities between different
vacuum configurations in the large-matrix limit. Despite these challenges,
substantial progress has been made through Monte Carlo simulations and related
algorithms~\cite{Delgadillo-Blando:2007mqd,Kovacik:2022kfh}.

In this work, we aim to better understand the implications of noncommutative
structure on phase transitions in matrix models. For this purpose, we analyse a field theory matrix model on
the fuzzy sphere~\cite{Madore:1991bw}, in which the kinetic term is taken to be
a cubic polynomial in the Laplacian operator.

In the large matrix-size limit, the eigenvalues of matrix models typically condense
into one or more disjoint intervals. The number and symmetry of these
intervals—referred to as the support of the eigenvalue distribution—determine
the phase of the system. In general, three phases are observed: the asymmetric
one-cut, symmetric one-cut, and symmetric two-cut phases, also known
respectively as the Ising, disordered, and non-uniform ordered phases. These
phases have been extensively studied in both commutative and noncommutative
settings~\cite{Brezin:1977sv,Shimamune:1981qf,Ambjorn:2002nj,Bietenholz:2004xs,Steinacker:2005wj,Steinacker:2005tf,OConnor:2007ibg,OConnor:2007pvc,GarciaFlores:2009hf,Ydri:2016daf,Ydri:2021cam}.

It is also known that, for matrix models with a kinetic term quadratic in the
Laplacian, these phases are separated by transitions forming a characteristic
phase diagram in which the three transition lines meet at a triple point, which
has been determined analytically and numerically~\cite{Polychronakos:2013fuzzy,Subjakova:2018iul,Subjakova:2020shi}. In our work, we study how the phase structure of the standard quadratic matrix model is modified upon introducing the cubic kinetic term, and whether the resulting model still exhibits a triple point. 

The paper is organised as follows. In Section~\ref{secc:2}, we review the basic ingredients necessary to formulate an effective matrix model. We discuss the $N=2$ cubic matrix model in Section~\ref{secc:3}, followed in Section~\ref{secc:4} by a detailed discussion of the different phases of the system. The analysis of phase transitions is presented in Section~\ref{secc:5}, where we determine numerically the location of the triple point as a function of the parameters of the model. We conclude with a discussion of our results. Technical details of some of the expressions used in the main text are collected in two appendices.

\section{Preliminaries on matrix models}
\label{secc:2}

Multitrace matrix models are known to sustain emergent geometries. In particular, the extended real quartic multitrace IKKT model reveals an emergent geometry on the fuzzy two-sphere.
At the heart of this model lies a random matrix theory with a uniform ordered phase.
Analysing the fluctuations around this configuration leads to a noncommutative $U(1)$ gauge theory coupled to a scalar field \cite{Ishibashi:1996xs}.

A dictionary must be established when formulating a scalar field theory on the fuzzy sphere. Fortunately enough, the passage from a scalar theory to the fuzzy domain is known: fields become Hermitian matrices, traces replace spatial integrals, and derivatives become inner derivatives \emph{i.e.}, commutators with fixed operators. In consequence, as a starting point, we consider the matrix model action
\be
S[\Phi] \equiv \mbox{Tr} \left[ \frac 1{2N^3} \Phi {\cal K}\Phi + \frac 1{2N} r \Phi^2 + V(\Phi) \right],
\label{matrixaction}
\ee
where $\Phi$ is a Hermitian $N\times N$ matrix representing a real scalar field, $V(\Phi)$ corresponds to the potential energy in the model whilst the term involving the operator ${\cal K}$ defines the kinetic term contribution.

The partition function of the model is the following integral over the spaces of $N \times N$ Hermitian matrices,
\begin{equation}
Z = \int d\Phi\, e^{-N^2 S[\Phi]}.
\end{equation}
Here, we adopt a normalization in which  $S[\Phi]$ and the eigenvalues of $\Phi$  scale as $N^0$.
This fixes the $N$-scaling of the various terms; $r$ is $N$-independent, $\mathrm{Tr}[\Phi\mathcal K\Phi]$ is of order $N^3$, and $\mathrm{Tr}[V(\Phi)]$ is of order $N^0$.

In the simplest case, ${\cal K}$ is nothing but the fuzzy version of the square angular momentum. In the case of the fuzzy sphere, it takes the form
\be
{\cal K} \Phi \equiv \frac1{R^2} \sum_{i=1}^3 [L_i, [L_i, \Phi]],
\label{eq:calKPhi}
\ee
where $L_i$ are the generators of $SU(2)$ in the $N$-dimensional representation, whilst $R$ represents the radius of the fuzzy sphere; usually $R=1$ for convenience. The sum of double commutators on the right-hand side of the above equation corresponds to the quadratic Casimir operator acting on $\Phi$; this operator is often denoted  $C_2$.
For the usual kinetic term, given by $\Tr(\Phi C_2 \Phi)$, the $N^3$ scaling follows from the aforementioned normalization, as we show later in this section. This motivates our choice of $N^3$ as the required scaling for the general kinetic term.

In the specific case of a noncommutative $\phi^4$-field theory, several approaches have been developed to analyse the theory defined by the action in Eq.~(\ref{matrixaction}).
When the kinetic term is absent, the model admits an exact solution due to its invariance under unitary conjugation $\Phi \to U \Phi U^\dagger$. Using the decomposition $\Phi = U \Lambda U^\dagger$, with $\Lambda = \mathrm{diag}(\lambda_1, \ldots, \lambda_N)$ the diagonal matrix of eigenvalues and $U$ a unitary matrix, the partition function can be written as an integral over the eigenvalues $\lambda_i$ and the angular degrees of freedom $U$. The integration over $U$ contributes only an overall constant, so that the effective model reduces to one depending only on the eigenvalues. However, this integration is no longer analytically possible when the kinetic terms are taken into account, due to the operators $L_i$. As a first step  in addressing this complication,  one can turn the $\phi^4$-field theory model into a multitrace matrix model by replacing the kinetic term by appropriate products of traces \cite{OConnor:2007ibg,OConnor:2007pvc}; it is known that the phase diagram of the resulting multitrace model shares several features with that of the original model.

Within this framework, an effective action for the eigenvalues can be found  in the
large-$N$ limit. For this purpose, one resorts to polarisation tensors
$\{\hat{Y}_{l, m}\}_{0 \leq l \le L,-l\leq m \le l}$, with $L = N - 1$,
which form a basis for the space of $N\times N$ matrices
and satisfy~\cite{Medina:2006},
\be
C_2 \hat Y_{l, m} \equiv  \sum_{i=1}^3 [L_i, [L_i, \hat Y_{l, m}]] = l(l +1) \hat Y_{l, m}, \qquad \hat Y_{l, m}^\dagger = (-1)^m \hat Y_{l,-m}\, .
\label{eq:c2ac}
\ee
When an additional scaling factor $\sqrt{4\pi/N}$ is incorporated into the $\hat Y_{l, m}$, we obtain the tensors $\hat T_{l, m}=\sqrt{4\pi/N} \hat Y_{l, m}$ satisfying the orthonormality condition $\mbox{Tr} (\hat T_{l, m}^\dagger \hat T_{l', m'}) = \delta_{ll'} \delta_{mm'}$ \cite{Varshalovich:1998}. By Eq.~(\ref{eq:c2ac}), the polarisation tensors $\hat Y_{l, m}$ are eigenvectors of the Casimir operator $C_2$. It follows that, for a general kinetic operator $\cal K$ depending solely on $C_2$, the following holds true
\be
{\cal K} \hat Y_{l, m} = \kappa[l(l + 1)] \hat Y_{l, m},
\label{kcasimir}
\ee
where the function $\kappa(x)$ defines the kinetic operator $\mathcal K=\kappa(C_2)$. In the simplest case of Eq.~(\ref{eq:calKPhi}), we have $\kappa(x) = x$, implying that $\mathcal K$ is, basically, the Casimir operator.

We see that, by expanding $\Phi$ in the basis of polarisation tensors
  $\hat Y_{l, m}$, the action of the Laplacian $C_2$ becomes diagonal, so that
  the kinetic term can be evaluated in terms of its eigenvalues $\kappa[l(l+1)]$.
Starting from this expression, if one writes $\Phi = U\Lambda U^\dagger$ and
  then performs the integration over the angular degrees of freedom, one obtains an effective action,
\begin{equation}
\int dU \, e^{-N^2 \bigl( \frac{1}{2N^3}\mathrm{Tr}[U\Lambda U^\dagger \mathcal K(U\Lambda U^\dagger) ] \bigr)}
= e^{-N^2 S_{\mathrm{eff}}(\Lambda)}.
\end{equation}
Although this integration cannot, in general, be carried out exactly,
in the large-$N$ limit, an approximate expression for $S_\mathrm{eff}$ can be derived \cite{Subjakova:2020shi}.
Specifically,  we can write $e^{- N^2 S_{\mathrm{eff}}(\Lambda)} \approx e^{-N^2F(t)/2}$,
with $t = \mathrm{Tr}[\big(\Lambda - \frac{1}{N}\mathrm{Tr}\Lambda\big)^2]/N$,
and  $F$  may be obtained after integrating the fundamental relation
\be
F'(t) = \frac 1t - f^{-1}(t)\, ,
\label{key}
\ee
where $f^{-1}(t)$ is the inverse function of
\be
f(z) \equiv \int_0^1 \frac {dx}{K(x) + z},
\label{functf}
\ee
and $K(x)=\lim_{N\to\infty} \kappa(N^2 x)/N^2$. The existence of this limit is equivalent to the $N^3$ scaling of the kinetic term, as shown in Eq.~\eqref{eq:kinetic_scaling} below.
 Evaluation of this integral in closed form is  possible for at least two particular choices of $K(x)$, namely, the linear and quadratic cases. Excluding these two situations, one usually resorts to a perturbative treatment of the integral according to the cases $1 \ll z$ or $z \ll 1$. The former leads to the integration of positive powers of $K(x)$, and the resulting expression for $f(z)$ is given by an infinite power series on $z^{-1}$ whose coefficients are straightforward to calculate. On the other hand, when $z \ll 1$, a careful evaluation of the integral is in order because the assumption $z \ll x$ fails near $x=0$.
To treat this regime, we employ the method of asymptotic matching \cite{Bender:1978},  details are given in Appendix~\ref{appendixa}.
 Once the asymptotic expressions for $f(z)$ are obtained, they can be reformulated in terms of $t=f(z)$, allowing one to determine $F(t)$ in the small- and large-$t$ regimes from Eq.(\ref{key}).

 As discussed in~\cite{Subjakova:2020shi}, the expression for  $F$ in the large-$t$ limit ($z \ll 1$) involves an integration constant $C$
--- see Eq.~(\ref{bigf}) ---
that is fixed by the condition
$F(0)=0$ and therefore cannot be determined from the $t \gg 1$ regime
alone. Determining the value of $C$ is crucial since
it appears in the large-$t$ expression for $F$ and must be fixed in order to consistently compare the free energies of the different phases
of the model.
In Appendix~\ref{appendixb}, we discuss the computation of the constant $C$ for the model under consideration.

Moreover, it has been argued that the presence of UV/IR mixing in matrix models derived from noncommutative $\phi^4$-field theory, is closely related to the existence of a two-cut phase. Input from noncommutative field theory points to the removal of this feature through a modified kinetic operator $\mathcal K$ involving powers higher than two of $C_2$; this kind of procedure is known as wave function regularization, and is used to suppress tadpole diagrams, which are responsible in part for this mixing. Such modifications of the kinetic term affect the location of the triple point in the models, leading to a reduction of the region where a two-cut phase is present. To understand this effect in more detail and to extend previous analyses, we consider in the following, a higher derivative cubic kinetic model defined by the action
\begin{equation}
S[\Phi] \equiv \mathrm{Tr}\left(\frac{1}{2N^3} (\Phi C_{2}\Phi + b \Phi C_{2}^{2}\Phi + c \Phi C_{2}^{3}\Phi) + \frac{r}{2N}\Phi^{2} + \frac{g}{4N}\Phi^{4}\right),
\end{equation}
which corresponds to $\kappa(x)=x+bx^2+cx^3$ and $V(\Phi)=g \Phi^4/4N$.
To preserve the $N^3$ scaling assumed in Eq.~(\ref{matrixaction}), the coefficients $b$ and $c$ may depend on $N$; the precise scaling required is discussed below.
Here we have assumed, without loss of generality, that the linear coefficient of $\kappa(x)$ is one. This can always be achieved by a rescaling of the field $\Phi$, together with a redefinition of the parameters $b$, $c$, $r$ and $g$.

After integrating out the angular degrees of freedom, we obtain an effective action for the eigenvalues,
\begin{equation}
  \int d\Phi e^{-N^2 S[\Phi]} \approx \int d\Lambda e^{ -N^2 \mathcal F[\Lambda]},\quad
\mathcal F[\Lambda]
= \frac{1}{2}F(t)
+ \frac{r}{2} c_2
+ \frac{g}{4} c_4
-\frac{1}{N^2} \sum_{i\neq j} \ln|\lambda_i - \lambda_j| ,
\label{eq:FLambda}
\end{equation}
where $c_n = \sum_i \lambda_i^n/N$, $t=c_2-c_1^2$, and the logarithmic term arises from the Jacobian associated with the diagonalization $\Phi = U \Lambda U^\dagger$.
In the large-$N$ limit, $\Lambda$ can be described by a spectral density $\rho$, and the partition function is evaluated by a saddle-point approximation,
$
Z \simeq e^{-N^2 \mathcal F[\rho_\star]},
$
where $\rho_\star$ minimizes $\mathcal F[\rho]$. The local minima of $\mathcal F$ correspond to the different phases of the theory:
the asymmetric one-cut, symmetric one-cut, and symmetric two-cut phases.

Each phase is characterized by a distinct support of the eigenvalue density.
With a slight abuse of notation, we denote the support parameters in the different phases by the same symbols $D$ and $\delta$.
The asymmetric one-cut, symmetric one-cut, and symmetric two-cut phases correspond respectively to
\begin{equation}
(D-\sqrt{\delta},\, D+\sqrt{\delta}), \qquad
(-\sqrt{\delta},\, \sqrt{\delta}), \qquad
\left(-\sqrt{D+\delta},\, -\sqrt{D-\delta}\right)\ \cup\ \left(\sqrt{D-\delta},\, \sqrt{D+\delta}\right).
\end{equation}
Within each phase, one assumes the corresponding support for the eigenvalue density and solves the saddle-point equations. This yields equations for the support parameters. Once these parameters are known, the moments $c_n$ and the corresponding free energy $\mathcal F$ of each phase can be computed \cite{Subjakova:2020shi}. We use these results in the following sections.

When $\kappa(x)$ is linear or quadratic in $x$, the critical lines of these matrix models have been determined, and analytic expressions are known for the constant $C$ in Eq.~(\ref{bigf}) ~\cite{Subjakova:2020shi}. Fortunately, although  more involved, one can also determine a closed expression for  $C$ in the cubic kinetic scenario as shown in the Appendix~\ref{appendixb}.  As with the quadratic matrix model studied in \cite{Subjakova:2018iul,Subjakova:2020shi} with $c=0$, these results lead to the analysis of phase transitions among asymmetric one-cut, symmetric two-cut and symmetric one-cut phases, as we discuss in detail in the next sections. Since it has been argued~\cite{OConnor:2007ibg,Kovacik:2018thy} that the multitrace model of the noncommutative $\phi^4$-field theory also exhibits these three phases, it is interesting to know how their properties change as we consider deviations from the standard kinetic term.

Some insights into  the expected behaviour for the triple point can be gained by considering
the contributions of the
different terms in Eq.~(\ref{eq:FLambda}).
When $r<0$, the term $r c_2/2$ repels eigenvalues from the origin; decreasing $r$ therefore favors the symmetric two-cut phase
over the symmetric one-cut phase.
The term $gc_4/4$ attracts the
eigenvalues to the origin, so increasing $g$ favors the symmetric one-cut phase.
The logarithmic term produces eigenvalue repulsion, preventing all
eigenvalues from collapsing into one. Finally, we consider the term involving $F$, which
encodes the contribution  of the kinetic operator to the eigenvalue distribution.
Expanding $\Phi = \sqrt{4\pi/N}\sum_{l, m}  \phi_{lm} \hat Y_{l, m}$, we find
$\Tr( \Phi \mathcal K \Phi)=\sum_{l, m} | \phi_{lm}|^2 \kappa[l(l+1) ]$, so increasing $c$ raises the kinetic cost of modes with $l>0$,
and configurations dominated by the $l=0$ mode become relatively favoured.
The $l=0$ mode corresponds to a multiple of the identity matrix, for which all eigenvalues coincide.
Therefore, increasing $c$ tends to favour the asymmetric one-cut phase.
As a result,  the triple point is expected to move towards more negative values of $r$ and larger values of $g$ as $c$ grows, so that the free energies of the competing phases coincide.

To further support the previous heuristic argument, we perform a high-temperature expansion, i.e., we treat the kinetic term perturbatively.
Introducing a parameter $a\ll1$ through $\kappa(x)=a(x+bx^2+cx^3)$, we can write,
\begin{equation}
  \begin{split}
\int dU \, e^{-N^2 \bigl( \frac{1}{2N^3}\mathrm{Tr}[U\Lambda U^\dagger \mathcal K(U\Lambda U^\dagger) ] \bigr)}
    =
\int dU \, \Bigl( 1-  N^2 \times \frac{1}{2N^3}\mathrm{Tr}[U\Lambda U^\dagger \mathcal K(U\Lambda U^\dagger) ]
\Bigr)
+O(a^2)
\,.
  \end{split}
\end{equation}
Using results from representation theory \cite{OConnor:2007ibg},
\begin{equation}
  \begin{split}
    \frac{1}{2N^3}
\int dU \,
\mathrm{Tr}[U\Lambda U^\dagger \mathcal K(U\Lambda U^\dagger) ]
=
\frac{\mathrm{Tr}[\mathcal K]}{2N^4} t+ O(1/N)
  \end{split}
\end{equation}
where $\Tr[\mathcal K]$
denotes the trace of $\mathcal K$ when viewed as an operator acting on the space of Hermitian matrices.
After exponentiating the previous result,
\begin{equation}
\int dU \, e^{-N^2 \bigl( \frac{1}{2N^3}\mathrm{Tr}[U\Lambda U^\dagger \mathcal K(U\Lambda U^\dagger) ] \bigr)}
= e^{-N^2 \frac{\mathrm{Tr}[\mathcal K]}{2N^4} t}+O(a^2)\,.
\end{equation}
By computing the trace using the basis of polarization tensors, in the $N\gg1$ limit we have,
\begin{equation}
  \begin{split}
\mathrm{Tr}[\mathcal K]&= \frac{4\pi}{N}\sum_{lm} \Tr(\hat Y_{l, m}^\dagger \mathcal K \hat Y_{l, m} )=
\sum_{lm} \kappa[ l(l+1) ] =\sum_{l=0}^{N-1} (2l+1) \kappa[ l(l+1) ] \sim
 \sum_{l=0}^{N-1}2 l \kappa( l^2) \sim \int_0^{N} dl\, 2 l \kappa( l^2)
                         \\&=
N^2 \int_0^1 dx\, \kappa( N^2 x) =N^4 \int_0^1 dx\, \frac{\kappa( N^2 x)}{N^2} \sim
N^4 \int_0^1 dx\, K( x)=
N^4\, \frac{a}{2} \Bigl( 1+\frac{2}{3}b+\frac{1}{2}c \Bigr)\, .
  \end{split}
\end{equation}
Within this approximation, the modified kinetic contribution is therefore
equivalent to an effective rescaling
$a \rightarrow a \Bigl( 1+\frac{2}{3}b+\frac{1}{2}c \Bigr)$. Since increasing $a$ moves the
triple point away from the origin of the $(g,r)$ plane \cite{OConnor:2007ibg},
this result provides further support for the expectation that increasing $c$
shifts the triple point towards larger $g$ and more negative $r$.

Finally, we establish the conditions under which the kinetic term scales as $N^3$, and show that this is indeed the scaling of the usual kinetic term in standard matrix models. On the one hand, from the definition of the polarization tensors, it follows that
\begin{equation}
  \Tr(\Phi \mathcal K \Phi)
  = \sum_{lm} |\phi_{lm}|^2 \kappa[ l(l+1) ]
  = \sum_{l} |\phi_{l}|^2 \kappa[l(l+1) ],
\end{equation}
where we have defined $|\phi_{l}|^2= \sum_{m} |\phi_{lm}|^2$. On the other hand,
since the eigenvalues of $\Phi$ are of order $N^0$, $\Tr(\Phi^2)=\sum_i | \lambda_i |^2$ is of
order $N$. However, this trace can also be expressed as
\[
\Tr( \Phi^2)= \sum_{lm} |\phi_{lm}|^2
= \sum_{l=0}^{N-1} |\phi_{l}|^2.
\]
Thus, as a heuristic large-$N$ estimate, assuming that the contributions $|\phi_l|^2$ do not exhibit a strong dependence on $l$, we take them to be of order $N^0$. Therefore, the leading behaviour of the kinetic term can be estimated as
\begin{equation}
\sum_{l} |\phi_{l}|^2 \kappa[ l(l+1) ]
\sim \sum_{l=0}^{N-1} \kappa[ l(l+1) ]
\sim \sum_{l=0}^{N-1} \kappa(l^2)
\sim \int_0^{N} dl\, \kappa(l^2).
\end{equation}
Suppose that $\kappa$ is a sum of terms of the form $A_\alpha x^\alpha$. The contribution of each term then scales as
\begin{equation}
A_\alpha \int_0^{N} dl\, l^{2\alpha}
= A_\alpha \frac{N^{2\alpha+1}}{2\alpha+1}.
\label{eq:kinetic_scaling}
\end{equation}
Therefore, obtaining the $N^3$ scaling requires
$A_\alpha = N^{2-2\alpha}\tilde A_\alpha$, with $\tilde A_\alpha$ independent of $N$.
This is precisely the behavior of $A_\alpha$ implied by requiring
$\lim_{N \rightarrow \infty} A_\alpha (N^2 x)^\alpha/N^2$
to exist. Thus, these two requirements are equivalent.

Note that for the usual kinetic term, $\kappa(x)=x$, the coefficient of the linear term requires no $N$-dependent rescaling, and hence $\Tr(\Phi C_2\Phi)$ already scales as $N^3$. More generally, if
$
\kappa(x)= \tilde a x +N^{-2}\tilde b x^2 +N^{-4}\tilde c x^3,
$
then
$
K(x)=\tilde a x+\tilde b x^2+\tilde c x^3,
$
where a tilde indicates a quantity independent of $N$. Henceforth, we drop the tildes to simplify the notation.

\section{A toy model: $N=2$ cubic matrix model}
\label{secc:3}

As pointed out in~\cite{GarciaFlores:2009hf}, the particular case of $N=2$ Hermitian matrices provides a way to understand phase transitions in field theory matrix models due to their complete integrability. Let us then consider the field
\begin{equation}
\Phi = \alpha {\mathbbold{1}}_{2\times 2} + \vec \rho \cdot \vec \sigma,
\end{equation}
where $\alpha$ is a real parameter and $\vec \rho$ is a three-dimensional vector with real components; here $\vec \sigma$ is the three-dimensional vector whose entries are the Pauli matrices $\{ \sigma_x, \sigma_y, \sigma_z \}$. Exploiting now the fact that
\begin{equation}
C_2 \Phi= \sum_{i=1}^3 [L_i, [L_i, \Phi]] =2 \vec \rho \cdot \vec \sigma,
\end{equation}
where $L_i, i = 1, 2, 3$ are the generators of $su(2)$ of dimension 2, we have
\begin{equation}
\mbox{Tr}[\Phi C_2 \Phi ] =\frac{1}{2} \mbox{Tr}[\Phi C_2^2 \Phi ] =\frac{1}{4} \mbox{Tr}[\Phi C_2^3 \Phi ] =4\rho^2,
\end{equation}
where $\rho \equiv |\vec \rho|$. Therefore, the action of our cubic matrix model, including the mass and interacting contributions,
\begin{equation}
S[\Phi]=\Tr \Bigl[a \Bigl(\Phi C_2 \Phi+\frac{b}{2} \Phi C_2^2 \Phi+\frac{c}{4}\Phi C_2^3 \Phi\Bigr)+r \Phi^2+g\Phi^4 \Bigr]
\end{equation}
 becomes explicitly
\begin{equation}
S[\alpha, \rho] = 4a \rho^2 (1 + b +c) +2r(\alpha^2 + \rho^2) + 2g(\alpha^4 + 6\alpha^2 \rho^2 + \rho^4).
\end{equation}
In the previous expression, we have included a scaling factor $a$ in front of the full kinetic term and rescaled the mass and coupling constants as well. 

\begin{figure}[!htb]
\includegraphics[width=0.7\textwidth]{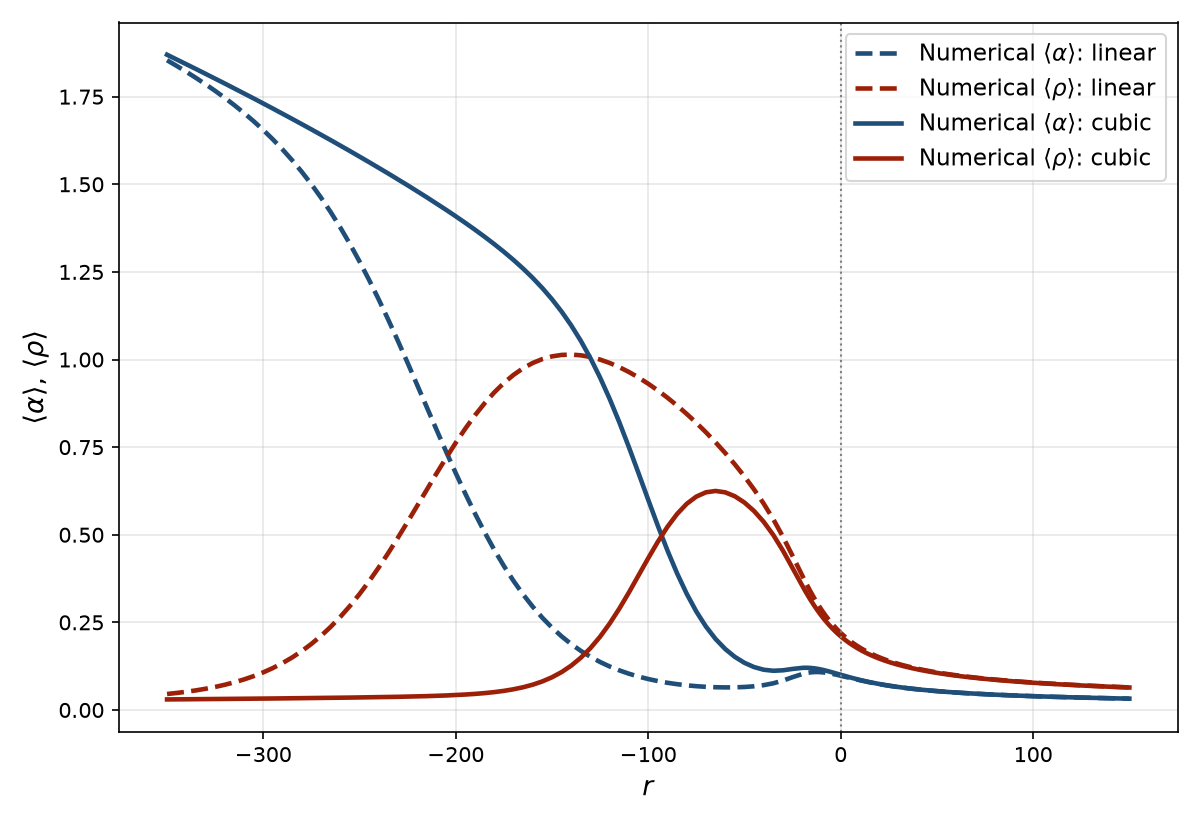}
\caption{Plot of the average values of $\alpha$ and $\rho$ in the $N=2$ linear (blue and red dashed lines respectively) and cubic (blue and red solid line respectively) matrix models as functions of the mass parameter $r$. The values of $b, c$ are $0.5, 0.3$ respectively; we also set $a=1, g = 50$. Notice that the uniform ordered phase in the cubic matrix model exists now for values of $r$ closer to zero, whilst the non-uniform ordered phase gets its range of existence reduced. The disordered phase remains practically unchanged.}
\centering
\label{averagealpharho}
\end{figure}

Once we have the action $S[\alpha,\rho]$, we can readily compute the expectation values $\langle \alpha \rangle$ and $\langle \rho \rangle$ to identify the different phases in the model. In Fig.~\ref{averagealpharho}, we show the results of the numerical evaluation of these quantities compared with the linear case~\cite{GarciaFlores:2009hf}. It is clear that the uniform ordered phase ($\alpha$-dominated) now extends its range of validity from $(-\infty, -300]$ to $(-\infty, -150]$ by taking its toll on the non-uniform ordered phase. Notice also that the values of $\langle \rho \rangle$ decrease in this phase for the cubic model. On the other hand, the disordered phase has not visible changes, giving vanishing expectation values for $\alpha$ and $\rho$ as $r$ increases. 

\begin{figure}
    \centering
    \subfigure[]{\includegraphics[width=0.31\textwidth]{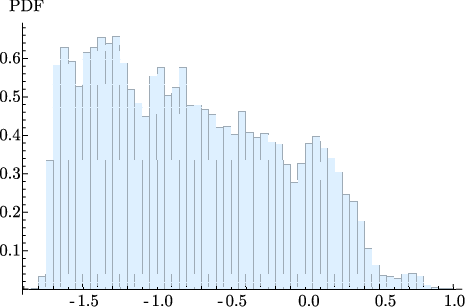}} 
    \subfigure[]{\includegraphics[width=0.31\textwidth]{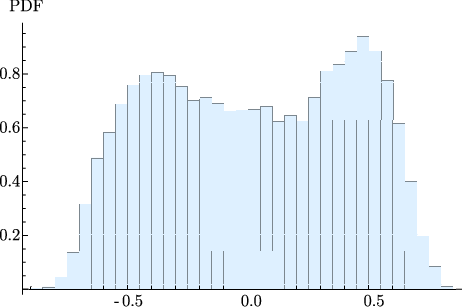}} 
    \subfigure[]{\includegraphics[width=0.36\textwidth]{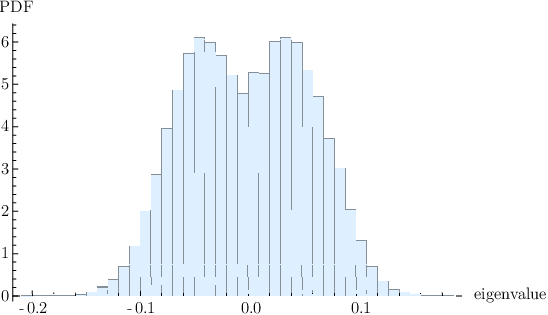}}
        \caption{Probability density functions for eigenvalues of the $N=2$ cubic matrix model from Monte Carlo simulations. The histograms illustrate the (a) uniform (one cut-asymmetric), (b) non-uniform (two cut-asymmetric) and (c) disordered (one cut-symmetric) phases with $r=-300, -50$ and $300$ respectively. The remaining parameters are fixed as $(a, b, c, g)=(1, 0.5, 0.3, 50)$.}
    \label{metromcn2}
\end{figure}

By using the Metropolis algorithm, we can also compute the probability density function of the eigenvalues for each phase. The histograms for the uniform, non-uniform and disordered phases are shown in Fig.~\ref{metromcn2}; the selected values of the mass parameter $r$ used to illustrate the phases are $-300, -50$ and $300$ respectively. 

\section{Phase transitions of the cubic Casimir kinetic matrix model}
\label{secc:4}

In this section, we analyse in detail the phase transitions of the cubic Casimir kinetic matrix model on the fuzzy sphere. Our model is based on the choice $K(x) \equiv x + bx^2 + cx^3$, meaning that we are dealing with a matrix model where the kinetic term is cubic, according to Eq.~(\ref{kcasimir}), in the Casimir invariant. Following~\cite{Subjakova:2020shi}, we calculate the function $F$ for our model, which has the form
\begin{align}
F(t) &\stackrel{t\to 0}{=} A_1 t + A_2 t^2 + A_3 t^3 + \dots,
\nonumber\\[4pt]
&\stackrel{t\to \infty}{=} \ln t + C + D_1 e^{-t} + D_2 e^{-2t} + D_3 e^{-3t} + \dots,
\label{bigf}
\end{align}
for the small and large-$t$ limits, respectively. Here the coefficients $A_i, i = 1, 2, \dots$, depend on the parameters $b$ and $c$, whilst $D_i, i =1, 2, \dots$, are polynomials of order $i-1$ in  $t$ with coefficients also depending on $b$ and $c$ (see Appendix~\ref{appendixa}). $C$ is a constant which has the explicit expression (see Appendix~\ref{appendixb})
\be
C \equiv \frac {b + 2c}{2c} \ln (1 + b+ c) - \frac {\sqrt{b^2 - 4c}}c \arccoth \left[ \frac {b + 2}{\sqrt{b^2 - 4c}} \right] - 2 = \frac {b + 2c}{2c} \ln (1 + b+ c) - \frac {\sqrt{b^2 - 4c}}{2c} \ln \frac {b + 2 + \sqrt{b^2 - 4c}}{b + 2 - \sqrt{b^2 - 4c}} - 2.
\label{constc}
\ee
The second form is particularly useful to take the limit $c \to 0$; we have in this situation
\be
C = \ln (1 + b) + \frac 1b \ln (1+b) - 1 + O(c),
\ee
that corresponds to the expression reported in~\cite{Subjakova:2020shi} for the quadratic kinetic matrix model. We now discuss the different phase transitions present in the cubic model.
All expressions for the free energy and the equations determining the support of the eigenvalue distributions are taken from~\cite{Subjakova:2020shi}.

\subsection{Large-$t$ expansion of the free energy along the transition line between symmetric phases}

Within the symmetric one-cut phase, the equation defining $\delta$ is
\be
\frac {4 - 3\delta^2 g}\delta - r - F'\left( t \right) = 0,
\label{eq:Fs1crel}
\ee where $t = (4\delta + \delta^3g)/16$. The previous equation determines $\delta$ for
arbitrary $g$ and $r$. However, since we are interested in the location of the
triple point, we evaluate it on the transition between the symmetric phases. On
this line one has $\delta = 2/\sqrt{g}$, independently of $F$
\cite{Polychronakos:2013fuzzy}; hence $t = 1/\sqrt{g}$. We can then solve
(\ref{eq:Fs1crel}) to parametrize the transition line, expressing $r$ as a
function of $g$. Notice that $\delta$ and $t$ are independent of $r$.

By employing the second line in Eq.~(\ref{bigf}), for small $g$, the resulting expression for $r$ has the general structure
\be
r = -5 \sqrt{g} + D_1 e^{-1/\sqrt{g}} + 2D_2 e^{-2/\sqrt{g}} + 3D_3 e^{-3/\sqrt{g}} + \dots.
\label{rs1c}
\ee
Furthermore, in the limit of small $g$, the free energy can be written in the following form,
\begin{equation}
  \mathcal{F}=
\frac {11}8 + \frac 12 \ln 2 + \frac 12 C + \frac {1}{2 } \frac r{\sqrt g} + \frac 12 \left( D_1 e^{-1/\sqrt g} + D_2 e^{-2/\sqrt g} + D_3 e^{-3/\sqrt g} + D_4 e^{-4/\sqrt g} + \dots \right),
\end{equation}
or, alternatively, using Eq.~(\ref{rs1c}), as
\begin{equation}
  \mathcal{F}=
-\frac 98 + \frac 12 \ln 2 + \frac 12 C + \frac 12 \left[ D_1 (1+ \sqrt g) e^{-1/\sqrt g} + D_2 (1 + 2\sqrt g) e^{-2/\sqrt g} + D_3 (1 +3 \sqrt g) e^{-3/\sqrt g} + \dots \right].
\label{eq:F1symsym}
\end{equation}

For the symmetric two-cut phase, the equation that determines $D$ is
\be
4 D g + r + F'(t) = 0,
\label{eq:EqDSymm}
\ee
where $t=D$ and $\delta=1/\sqrt{g}$.
The transition to the symmetric one-cut phase occurs at
$D = \delta= 1/\sqrt{g}$. Note that the $\delta$ appearing here corresponds to the support of the symmetric two-cut phase, and is distinct from that of the symmetric one-cut phase in (\ref{eq:Fs1crel}).
Solving this equation for $r$ yields the same result as in the symmetric one-cut case, namely Eq.~(\ref{rs1c}).
Consistently, a direct calculation also shows that the corresponding free energy coincides with
(\ref{eq:F1symsym}).

\subsection{Small-$t$ expansion of the free energy in the asymmetric one-cut phase}

In this case, the equation that determines $\delta$
is,
\be
 \frac {4(4 + 15 \delta^2 g + 2 r\delta)}{\delta(4 + 9\delta^2 g)} - F'(t) = 0,
\label{eq:FAsymmcut1}
\ee
with $t = \delta(64+ 160 \delta^2 g + 144 \delta^4 g^2 + 81 \delta^6 g^3 + 36 \delta^3 gr + 27 \delta^5 g^2 r)/64(4 + 9\delta^2 g)$. The equation that
determines $D$ is not needed in our analysis and it is not included here. To solve (\ref{eq:FAsymmcut1}) for $\delta$,
we focus on the limit $r\rightarrow -\infty$. In this regime, the logarithmic repulsion term in Eq.~(\ref{eq:FLambda}) becomes irrelevant and all the eigenvalues tend to collapse to the same point, so $\delta$ goes to zero. We therefore seek a solution of Eq.~(\ref{eq:FAsymmcut1}) as an expansion in inverse powers of $r$,
\be
\delta = \frac {\delta_1}r + \frac {\delta_2}{r^2} + \frac {\delta_3}{r^3} + \dots
\ee
where the coefficients $\delta_i, i=1, 2, \dots$ are polynomials of degree $\lceil i/2 \rceil - 1$  on $g$. Here $\lceil n \rceil$ denotes the ceiling function applied to $n$; we have, for example, $\delta_1 = -2$. The general structure of $\mathcal F$ then becomes
\be
\mathcal F=
-\frac {r^2}{16g} + \frac 12 \ln (-r) + \frac 34 + \frac 12\ln 2 + \frac {f_1}r + \frac {f_2}{r^2} + \dots,
\label{fa1c}
\ee
where $f_i, i=1, \dots$, is a polynomial of order $\lceil (i-1)/2 \rceil $ on $g$.

\subsection{Large-$t$ expansion of the symmetric two-cut free energy neglecting exponentially suppressed terms}

For large-$t$, Eq.~(\ref{bigf}) shows that the exponential contributions to $F(t)$ are heavily suppressed; therefore, we may work with the simple form $F(t) = \ln t + C$.
Within this approximation,
 to solve (\ref{eq:EqDSymm}) for $D$, we look for a series expansion
\be
D = \alpha r + d_0 + \frac {d_1}r + \frac {d_2}{r^2} + \dots,
\ee
in inverse powers of $r$. It follows that the only non-vanishing coefficients $d_i$ are those with $i$ odd, leading to the expression
\be
D = -\frac r{4g} + \frac 1r + \frac {4g}{r^3} + \frac {32 g^2}{r^5} + \frac {320 g^3}{r^7} + \dots\,.
\label{ds2clarget}
\ee
Using the above result, the free energy is expressed as the series expansion
\be
\mathcal F=
- \frac {r^2}{16g} + \frac 12 \ln (-r) + \frac 38 + \frac 12 C - \frac 14 \ln(4g) - \frac g{r^2} - \frac {4g^2}{r^4} - \frac {80g^3}{3r^6} - \frac {224g^4}{r^8} + \dots,
\label{fs2cnonexp}
\ee
where the constant $C$ is defined in Eq.~(\ref{constc}). At this point, it is worthwhile to remark that due to the simplicity of the function $F(t)$ for this case, a closed form for $D$ and the free energy can be found. To see this, we simply substitute $F'(t)$ by
$1/t=1/D$ in Eq.~(\ref{eq:EqDSymm}) and solve the resulting quadratic equation for $D$ to obtain
\be
D^{\pm} = \frac {-r \pm \sqrt{r^2 - 16g}}{8g}.
\ee
It is straightforward to verify that when $r \to -\infty$, the solution $D^+ = (-r + \sqrt{r^2 - 16g})/8g$ reproduces the series expansion in Eq.~(\ref{ds2clarget}); furthermore, by calculating the free energy associated with $D^+$, we obtain
\be
\mathcal{F}=
\frac 18 + \frac 12 C - \frac {r^2(1 - \sqrt{1 - 16gr^{-2}})}{32g} + \frac 12 \ln(-r) + \frac 12 \ln \frac {1 - \sqrt{1 -16g r^{-2}}}2 - \frac 14 \ln(4g).
\label{exactfree1}
\ee
A series expansion of this expression when $r\to -\infty$ gives Eq.~(\ref{fs2cnonexp}) immediately.

\subsection{Large-$t$ expansion of the symmetric one-cut free energy neglecting exponentially suppressed terms}

In the $t\to +\infty$ limit, exponentially suppressed terms are again negligible, so we set $F(t) = \ln t + C$  and solve (\ref{eq:Fs1crel}) for $\delta$
by using the series expansion
\be
\delta = \delta_0 + \delta_1 (r + 5\sqrt{g}) + \delta_2 (r + 5\sqrt{g})^2 + \dots\, .
\ee
This Ansatz is motivated by the fact that
the triple point lies close to the transition between the symmetric phases,
which is located near $r+5\sqrt g=0$; see Eq.~(\ref{rs1c}).

The result is
\be
\delta = \frac 2{\sqrt g} - \frac {r + 5\sqrt{g}}{3g} - \frac {(r + 5\sqrt{g})^2}{216g^{3/2}} - \frac {7(r + 5\sqrt{g})^3}{7776g^2}  - \frac {137(r + 5\sqrt{g})^4}{1119744g^{5/2}} - \frac {565(r + 5\sqrt{g})^5}{40310784g^3} + \dots,
\label{deltas1clarget}
\ee
The free energy in powers of $r + 5\sqrt{g}$ is straightforwardly calculated as
\be
\mathcal F=
-\frac 98 + \frac 12 \ln 2 + \frac 12 C + \frac {r + 5\sqrt{g}}{2\sqrt{g}} - \frac {(r + 5\sqrt{g})^2}{12g} + \frac {(r + 5\sqrt{g})^3}{162g^{3/2}}  - \frac {(r + 5\sqrt{g})^4}{3888g^2} + \frac {(r + 5\sqrt{g})^5}{699840g^{5/2}} + \dots
\label{fs1cnonexp}
\ee
Notice that, as before, the free energy depends on the parameter $C$ only through the presence of the term $C/2$.

As in the previous subsection, the simple large-$t$ form of $F(t)$ allows one to write closed expressions for $\delta$ and the free energy. For this purpose, we notice that Eq.~(\ref{eq:Fs1crel}) is a cubic equation for $\delta$; the only real solution to it is given by
\be
\delta = -\frac{\Delta^{1/3}}{9 g^2}+\frac{72 g^3-g^2 r^2}{9 g^2 \Delta^{1/3}}-\frac{r}{9 g},
\label{deltacloseds1clarget}
\ee
where $\Delta \equiv g^3 r^3+18 \sqrt{3} \sqrt{131 g^8 r^2+g^7 r^4+384 g^9}+378 g^4 r$. It is not difficult to check that the series expansion of the above expression for $\delta$ around the value $r=-5\sqrt{g}$ coincides precisely with Eq.~(\ref{deltas1clarget}). On the other hand, it is also not hard to verify that, by employing Eq.~(\ref{deltacloseds1clarget}), the  free energy can be written as
\be
\mathcal F=
\frac{1}{4} + \frac{C}{2}-\frac{r \Omega }{72 g^2}-\frac{r \Omega ^3}{23328 g^5}+\frac{\Omega ^4}{93312 g^6}+\frac{1}{2} \ln \left[-\frac{\Omega
   \left(\frac{\Omega ^2}{81 g^3}+4\right)}{144 g^2}\right] - \frac{1}{2} \ln \left(-\frac{\Omega }{36 g^2}\right),
\label{exactfree2}
\ee
where $\Omega \equiv gr + g^2(r^2 - 72g)\Delta^{-1/3} + \Delta^{1/3}$. As before, a series expansion of this result around $r=-5\sqrt{g}$ gives the free energy as in Eq.~(\ref{fs1cnonexp}).

\section{Numerical analysis of the phase diagram of the cubic Casimir kinetic matrix model}
\label{secc:5}

The transition line between the symmetric one-cut and symmetric two-cut phases is straightforward to compute and is given by Eq.~(\ref{rs1c}). On the other hand, the phase transition between the asymmetric one-cut and symmetric two-cut phases requires matching the free energies of Eqs.~(\ref{fa1c}) and (\ref{fs2cnonexp}).
Following the procedure used for the quadratic kinetic matrix model in
\cite{Subjakova:2020shi}, we proceed by calculating the difference between these two equations, setting it equal to zero and solving for $g$ as a function of $r$ by assuming
\be
g = g_0 + \frac {g_1} r + \frac {g_2}{r^2} + \dots .
\label{grequ}
\ee
This ansatz gives a system of coupled algebraic equations for the unknown coefficients $g_i, i= 1,2, \dots$; proceeding along this way, we obtain, for example, $g_0 = \exp(2C - 3/2 - 4 \ln 2)$. Since Eq.~(\ref{grequ}) is only valid for large $r$, we need to extrapolate it if contact with small values of $r$ is sought, namely, if we are near the origin in the $(g, r)$-plane. For this purpose, Padé approximants are useful with the constraint that as $r \to 0$, we must also have $g \to 0$.

The transition from asymmetric to symmetric one-cut phase is more technically involved than the previous phase transitions, similar to the situation encountered in the quadratic kinetic matrix model  studied in \cite{Subjakova:2020shi} (where $c=0$). In particular, we need to rewrite Eq.~(\ref{fa1c}) using Padé approximants in powers of $r+5\sqrt{g}$ to make a proper comparison with the series expansion in Eq.~(\ref{fs1cnonexp}). Nevertheless, the numerical implementation can be done straightforwardly, and thus, the phase transition line can be found accordingly. For our purposes, we employed a Padé approximant of order [9/9].

Once we have determined the transition lines, we  proceed to calculate the critical values of $g$ associated with the existence of a triple point. In Table~\ref{table1}, we present the numerical results for the critical value of $g$ obtained from the intersection between the asymmetric one-cut/symmetric two-cut transition line and the symmetric one-cut/symmetric two-cut transition line, as the parameters $b$ and $c$ of the model are varied. We denote this value by $g_c^{a1s2}$. For small values of $c$, the numerical results remain close to those reported for the quadratic model (with $c=0$) as the parameter $b$ is varied. However, as the value of the parameter $c$ increases, the values of $g_c^{a1s2}$ also increase, moving away from the critical values associated with the quadratic model.

\begin{table}[!htb]
\begin{center}
\caption{Numerical values for the critical value $g^{a1s2}_c$ obtained from
the intersection of the asymmetric one-cut/symmetric two-cut and symmetric one-cut/symmetric two-cut transition lines, for different values of the parameters $b$ and $c$.}
\begin{tabular}{|c|c|c|c|c|c|}\hline
\backslashbox{b}{c}
&\makebox[3em]{0.001}&\makebox[3em]{0.01}&\makebox[3em]{0.1}&\makebox[3em]{0.3}&\makebox[3em]{0.6}\\\hline\hline
\, -0.25 \, & \, 0.00388718 \, &\, 0.00390933 \, & \, 0.00412782 \, & \, 0.00459717 \, & \, 0.00527147 \, \\\hline
\, 0 \, & \, 0.00486816 \, &\, 0.00488997 \, & \, 0.00510604 \, & \, 0.00557480 \, & \, 0.00625584 \, \\\hline
\, 0.25 \, & \, 0.00588138 \, &\, 0.00590316 \, & \, 0.00611946 \, & \, 0.00659152 \, & \, 0.00728227 \, \\\hline
0.5 & \, 0.00693124 \,& \, 0.00695315 \,& \, 0.00717115 \, & \, 0.00764877 \, & \, 0.00835109 \, \\\hline
\end{tabular}
\label{table1}
\end{center}
\end{table}

A similar behaviour is observed for the critical values $g^{a1s1}_c$
associated with the intersection between the asymmetric one-cut/symmetric one-cut and symmetric one-cut/symmetric two-cut  transition lines. The corresponding numerical values are given in Table~\ref{table2}.
Consistent with the behaviour of $g_c^{a1s2}$, as the parameter $c$ increases, the critical values are shifted to larger values, showing an increase of around $60\%$ when $c=0.6$.

\begin{table}[!htb]
\begin{center}
\caption{Numerical values for the critical value $g^{a1s1}_c$, obtained from intersection of the asymmetric one-cut/symmetric one-cut and symmetric one-cut/symmetric two-cut transition lines.}
\begin{tabular}{|c|c|c|c|c|c|}\hline
\backslashbox{b}{c}
&\makebox[3em]{0.001}&\makebox[3em]{0.01}&\makebox[3em]{0.1}&\makebox[3em]{0.3}&\makebox[3em]{0.6}\\\hline\hline
\, -0.25 \, & \, 0.00388526 \, &\, 0.00390751 \, & \, 0.00412652 \, & \, 0.00459615 \, & \, 0.00527048 \, \\\hline
\, 0 \, & \, 0.00486704 \, &\, 0.00488886 \, & \, 0.00510497 \, & \, 0.00557376 \, & \, 0.00625473 \, \\\hline
\, 0.25 \, & \, 0.00588028 \, &\, 0.00590206 \, & \, 0.00611835 \, & \, 0.00659037 \, & \, 0.00728103 \, \\\hline
0.5 & \, 0.00693004 \,& \, 0.00695196 \,& \, 0.00716993 \, & \, 0.00764748 \, & \, 0.00834970 \, \\\hline
\end{tabular}
\label{table2}
\end{center}
\end{table}

\begin{figure}[!htb]
\includegraphics[width=0.6\textwidth]{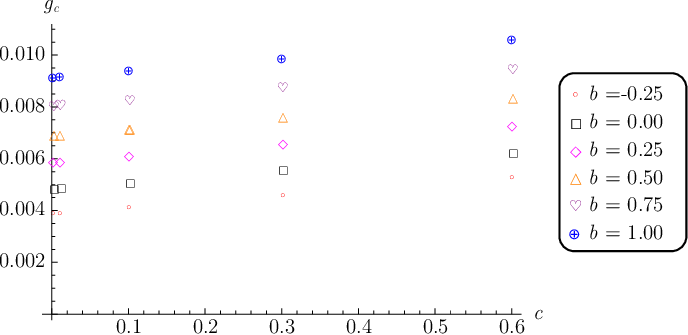}
\caption{Plot of the critical value $g_c$ as a function of the parameters $b$ and $c$ for the cubic matrix model. The values employed for $c$ are $0.001, 0.01, 0.1, 0.3$ and $0.6$ (left to right). It is clear that as the parameter $c$ increases, for fixed $b$, the value of $g_c$ also increases. Furthermore, numerical evaluation shows that as $c\to 0$, we recover the value of $g_c$ associated with the quadratic matrix model.}
\centering
\label{figgcrit}
\end{figure}

We see that the values of the critical $g_c$ calculated numerically, using different phase transition lines, agree quite well --- up to five digits after the decimal point with a relative difference of around $0.02\%$--- pointing out to the existence of a triple point for the model. Furthermore, we have verified that the evaluation of the non-perturbative analytical expressions Eqs.~(\ref{exactfree1}) and~(\ref{exactfree2}) for the free energy of the symmetric one-cut and two-cut phases, respectively, leads to the same outcome when evaluated along the curve $r=-5\sqrt{g}$. Therefore, we infer the existence of a triple point where the three phases coexist. We also conclude from the preceding discussion that in the cubic kinetic matrix model, the critical value $g_c$  increases with the parameter $c$. This behaviour is better illustrated in Fig.~\ref{figgcrit}, where we plot the numerical values of $g_c$ as $c$ varies with $b$ fixed, using the data from Table~\ref{table1} together with other choices of the parameters $b$ and $c$.

\begin{figure}[!htb]
\includegraphics[width=0.50\textwidth]{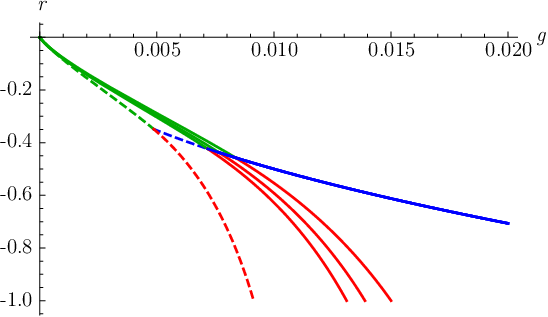}
\caption{Phase transition lines for the cubic Casimir matrix model.  The green, red and blue solid lines correspond to the asymmetric one-cut to symmetric one-cut, asymmetric one-cut to symmetric two-cut and symmetric one-cut to symmetric two-cut phase transitions respectively. Values of the parameters are $b=0.5$ and $c=0.1, 0.3$ and $0.6$ (from left to right); it is clear that the triple point gets displaced to the right as $c$ increases its value. Dashed lines correspond to the phase transition lines of the linear Casimir model ($b=0=c, g_c = 0.0048655$).} 
\centering
\label{figphaselines}
\end{figure}

Based on our previous analysis, we build the phase transition diagram shown in Fig.~\ref{figphaselines} for generic values $b=0.5, c=0.1, 0.3$ and $0.6$; with these choices, $g_c \approx 0.00717, 0.00765$ and $0.00835$ respectively. In this plot, following the colour code in~\cite{Subjakova:2020shi}, the symmetric one-cut/asymmetric one-cut phase transition line is shown in green, whilst the symmetric two-cut/asymmetric one-cut phase transition line is shown in red; the blue line corresponds to the remaining symmetric one-cut/symmetric two-cut transition line. For comparison, we also show in this plot the transition lines of the linear Casimir model where $g_c=0.0048655$. The location of the triple point is clearly identified, illustrating how its position depends on the parameter $c$, moving to the right as $c$ increases its value.
It is worth to mention that larger values of $g_c$ can be also be achieved by increasing $c$, even when $b$ is small or negative,  leading to a phase diagram where the triple point and the boundaries among the phase transitions are clearly identified. %
These findings are consistent with the heuristic arguments presented at the end of Section~\ref{secc:2} regarding the behaviour of the triple point.

\section{Conclusions}

The cubic kinetic matrix model allows us to discuss the influence of higher derivative terms in the free energy of matrix models over the fuzzy sphere. We have shown that the general features of matrix models on the fuzzy sphere remain; we find
the asymmetric one-cut, symmetric one-cut, and symmetric two-cut phases (corresponding respectively to the uniform ordered, disordered, and non-uniform ordered phases),
and in this regard, our numerical analysis provides evidence the existence of a triple point where these phases coexist. An important point to draw attention to is that the location in the $(g,r)$-plane of the triple point in our model can be controlled by the parameters $b$ and $c$; we have given explicit expressions for the effective free energy of the symmetric one-cut and two-cut phases in the large-$t$ limit, such expressions agree along the curve $r=-5\sqrt{g}$.

Furthermore, we have determined the constant $C$
--- without resorting to numerical or perturbative methods ---
 necessary to completely determine the effective free energy in our model.  A similar procedure for other higher derivative models may also lead to explicit expressions of the corresponding constants. Extending this procedure to other higher-derivative kinetic models would therefore be an interesting direction for future work.

Our model represents a non-trivial extension of the standard matrix models on the fuzzy sphere beyond linear and quadratic derivative kinetic terms.
Our results confirm the expected behaviour of matrix models regarding the existence of a triple point, which is thought to be a generic feature in them. Furthermore, we have also shown that higher polynomial derivative terms in these models move the location of the triple point to large values of the coupling constant $g$~\cite{GarciaFlores:2009hf}. In particular, by suitably choosing the parameter $c$ that modulates the cubic kinetic term, we can shift the triple point to values of $g$ larger than the standard linear kinetic contribution, reducing the two-cut phase region. Such behaviour is commonly associated with a suppression of the UV/IR mixing in a matrix model defined over the fuzzy sphere; in our model, this mechanism is enhanced as long as the parameter $c$ remains positive.

\acknowledgments
A. T. acknowledges financial support from a SECIHTI PhD grant (CVU 266018).

\appendix

\section{$f(z)$ through asymptotic matching of integrals}
\label{appendixa}

Here, we outline how to evaluate Eq.~(\ref{functf}) for the two cases $1 \ll z$ and $z \ll 1$ when dealing with a cubic $K(x) \equiv x + bx^2 + cx^3$. For the former case, we proceed as follows
\begin{eqnarray}
\int_0^1 dx \frac 1{x + bx^2 + cx^3 + z} &=& \frac 1z \int_0^1 dx \frac 1{1 + \dfrac {x + bx^2 + cx^3}z} = \frac 1z \int_0^1 dx \left[ 1 - \frac {x + bx^2 + cx^3}z + \left( \frac {x + bx^2 + cx^3}z \right)^2 +\dots \right]
\nonumber \\[4pt]
&=& \frac 1z \left[1 - \frac {6+4b+3c}{12}\frac 1z + \left( \frac 13+\frac {b^2}5+\frac {2c}5 + \frac{c^2}7+ \frac {b(3+2c)}6 \right) \frac 1{z^2} + \dots \right].
\end{eqnarray}
Clearly, the coefficients in the resulting infinite power series in $z^{-1}$ can be obtained straightforwardly; they are polynomials on the parameters $b$ and $c$, and we need not large amounts of CPU time to calculate them. Let us now consider the situation where $z \ll 1$; we start by splitting the integral in Eq.~(\ref{functf}) as
\be
\int_0^1 dx \frac 1{x + bx^2 + cx^3 + z} = \int_0^\delta dx \frac 1{x + bx^2 + cx^3 + z} + \int_\delta^1 dx \frac 1{x + bx^2 + cx^3 + z},
\ee
where $\delta \ll 1$ is an auxiliary parameter separating the inner region (first integral with $x \ll 1$) from the outer region (second integral with $z \ll x$) of the integration interval~\cite{Bender:1978}; $\delta$ is such that $z \ll \delta \ll 1$ but at the end, the integral that we seek out to calculate should not depend on it.
In the outer region, we have
\begin{equation}
I^{out}=\int_\delta^1 dx\frac 1{x + bx^2 + cx^3 + z} = \sum_{n=0}^\infty (-z)^n \int_\delta^1 dx \frac 1{(x + bx^2 + cx^3)^{n+1}}
\end{equation}
We first compute,
\begin{eqnarray}
I_0^{out} \equiv \int_\delta^1 dx \frac 1{x + bx^2 + cx^3} &=& \int_\delta^1 dx \left\{ \frac 1x -\frac {b + 2cx}{1 + bx + cx^2} + \frac {cx}{1 + bx + cx^2} \right\}
\nonumber \\[4pt]
&=& \ln x - \ln(1 + bx + cx^2) \Big|_\delta^1 +  \int_\delta^1 dx \frac {cx}{1 + bx + cx^2}
\end{eqnarray}
We evaluate the last integral through a change of variable as follows
\begin{eqnarray}
\int_\delta^1 dx \frac {cx}{1 + bx + cx^2} & \stackrel{x \equiv s/c}{=} & \int_{c\delta}^c ds \frac s{c + bs + s^2} = \int_{c\delta}^c ds \frac s{(s - s_+)(s - s_-)} = \frac 1{s_+ - s_-} \ln \frac {(s - s_+)^{s_+}}{(s - s_-)^{s_-}} \Big|_{c\delta}^c
\nonumber \\[4pt]
&=& \frac 1{s_+ - s_-} \left [ \ln \frac {(c - s_+)^{s_+}}{(c - s_-)^{s_-}} - \ln \frac {(c\delta - s_+)^{s_+}}{(c\delta - s_-)^{s_-}} \right] =: w(c, 1) - w(c, \delta)
\end{eqnarray}
where $s_\pm \equiv (-b \pm \sqrt{b^2 - 4c})/2$. Notice that as $c \to 0$ we have $s_+ \to 0, s_- \to -b$ implying $s_+ - s_- \neq 0$; the integral thus vanishes in this limit. Here, we would like to point out that if we evaluate $I^{out}_0$ with Mathematica, we obtain the result
\be
\int_\delta^1 dx \frac {cx}{1 + bx + cx^2} = \ln x - \frac 12 \ln(1 + bx + cx^2) - \frac b{\sqrt{-b^2 + 4c}} \arctan \frac {b + 2cx}{\sqrt{-b^2 + 4c}} \Big|_\delta^1,
\ee
which, though in principle correct, is not suitable to discuss the limit $c\to 0$ because of the term involving $\arctan$. Indeed, the above result comes from expressing the integrand as
\be
\frac 1x - \frac 12 \frac {b+2cx}{1 + bx + cx^2} - \frac 12 \frac b{1 + bx + cx^2}, %
\ee
and if the last term is rewritten as
\be
\frac 12 \frac b{1 + bx + cx^2} = \frac {2bc}{-b^2 + 4c + (b + 2cx)^2} = \frac b{(-b^2 + 4c)} \frac {2c}{1 + \dfrac {(b + 2cx)^2}{-b^2 + 4c}},
\ee
we introduce a spurious singularity in the integrand of $I_0^{out}$ since the numerator and denominator of the previous expression vanish as $c \to 0$, whilst the original expression for the integrand is non-singular at $c=0$. For this reason, when evaluating this kind of integrals with Mathematica, the replacement
\be
\frac b{\sqrt{-b^2 + 4c}} \arctan \frac {b + 2cx}{\sqrt{-b^2 + 4c}} \Big|_\delta^1 \rightarrow \frac 12 \ln(1 + bx + cx^2) \Big|_\delta^1 - \frac 1{s_+ - s_-} \ln \frac {(s - s_+)^{s_+}}{(s - s_-)^{s_-}} \Big|_{c\delta}^c,
\ee
should be used to obtain expressions with a smooth limit $c\to 0$. With this in mind, for example, we have up to quadratic terms in $z$, the result
\begin{eqnarray}
  I^{out}=
&&\ln \left(b
   \delta +c \delta ^2+1\right)-\ln (b+c+1)-\ln (\delta )-w(c,\delta )+w(c,1) + z \left(-\frac{b^2 c \delta +b^3-3 b c-2 c^2 \delta }{\left(b^2-4 c\right)
   \left(b \delta +c \delta ^2+1\right)} \right.
   \nonumber \\[4pt]
   &&\left. +\frac{2 \left(-6 b^2 c+b^4+6
   c^2\right) \left(w(c,\delta )-\frac{1}{2} \ln \left(b \delta +c \delta
   ^2+1\right)\right)}{b \left(4 c-b^2\right)}+b \ln \left(b \delta +c
   \delta ^2+1\right)-2 b \ln (\delta )-\frac{1}{\delta }\right)
   \nonumber \\[4pt]
   &&
   -z
   \left(\frac{2 \left(-6 b^2 c+b^4+6 c^2\right) \left(w(c,1)-\frac{1}{2} \ln
   (b+c+1)\right)}{b \left(4 c-b^2\right)}-\frac{b^2 c+b^3-3 b c-2
   c^2}{\left(b^2-4 c\right) (b+c+1)}+b \ln (b+c+1)-1\right)
\nonumber \\[4pt]
&&
+\frac {z^2}2 \left( \frac{6 \left(70 b^2 c^2-21 b^4 c+2 b^6-70 c^3\right) \left(w(c,1)-\frac{1}{2} \ln
   (b+c+1)\right)}{\left(4 c-b^2\right)^2}+\frac{b^3 c-4 b^2 c+b^4-3 b c^2+2
   c^2}{\left(b^2-4 c\right) (b+c+1)^2} \right.
 \nonumber \\[4pt]
&&
\left.   +\frac{-42 b^3 c^2+97 b^2 c^2+6 b^5 c-47 b^4 c+6
   b^6+66 b c^3-32 c^3}{\left(b^2-4 c\right)^2 (b+c+1)}+3 \left(c-2 b^2\right) \ln
   (b+c+1)+6 b-1 \right)
   \nonumber \\[4pt]
&&
-\frac {z^2}2 \left( \frac{-42 b^3 c^2 \delta +97 b^2 c^2+6 b^5 c \delta -47 b^4 c+6 b^6+66 b c^3 \delta -32
   c^3}{\left(b^2-4 c\right)^2 \left(b \delta +c \delta ^2+1\right)}+\frac{b^3 c \delta
   -4 b^2 c+b^4-3 b c^2 \delta +2 c^2}{\left(b^2-4 c\right) \left(b \delta +c \delta
   ^2+1\right)^2}  \right.
 \nonumber \\[4pt]
&&
 +\frac{6 \left(70 b^2 c^2-21 b^4 c+2 b^6-70 c^3\right) \left(w(c, \delta
   )-\frac{1}{2} \ln \left(b \delta +c \delta ^2+1\right)\right)}{\left(4
   c-b^2\right)^2}+3 \left(c-2 b^2\right) \ln \left(b \delta +c \delta ^2+1\right)
 \nonumber \\[4pt]
&&
\left. +6
   \left(2 b^2-c\right) \ln (\delta )+\frac{6 b}{\delta }-\frac{1}{\delta ^2} \right)
   +O(z^3)
   \nonumber \\[4pt]
\end{eqnarray}
We now make a series expansion on $\delta$ up to linear terms for later convenience; we obtain
\begin{eqnarray}
I^{out}&=&  -\ln (b+c+1)-\ln (\delta ) -w(c,0 )+w(c,1) + z \left(-\frac{b^3-3 b c }{\left(b^2-4 c\right)} +\frac{2 \left(-6 b^2 c+b^4+6
   c^2\right) w(c, 0) }{b \left(4 c-b^2\right)} \right.
   \nonumber \\[4pt]
   &&\left.  - 2 b \ln (\delta )-\frac{1}{\delta }\right)
    \nonumber \\[4pt]
   &&
   -z
   \left(\frac{2 \left(-6 b^2 c+b^4+6 c^2\right) \left(w(c,1)-\frac{1}{2} \ln
   (b+c+1)\right)}{b \left(4 c-b^2\right)}-\frac{b^2 c+b^3-3 b c-2
   c^2}{\left(b^2-4 c\right) (b+c+1)}+b \ln (b+c+1)-1\right)
   \nonumber \\[4pt]
&&
+\frac {z^2}2 \left( \frac{6 \left(70 b^2 c^2-21 b^4 c+2 b^6-70 c^3\right) \left(w(c,1)-\frac{1}{2} \ln
   (b+c+1)\right)}{\left(4 c-b^2\right)^2}+\frac{b^3 c-4 b^2 c+b^4-3 b c^2+2
   c^2}{\left(b^2-4 c\right) (b+c+1)^2} \right.
 \nonumber \\[4pt]
&&
\left.   +\frac{-42 b^3 c^2+97 b^2 c^2+6 b^5 c-47 b^4 c+6
   b^6+66 b c^3-32 c^3}{\left(b^2-4 c\right)^2 (b+c+1)}+3 \left(c-2 b^2\right) \ln
   (b+c+1)+6 b-1 \right)
   \nonumber \\[4pt]
&&
-\frac {z^2}2 \left( \frac{97 b^2 c^2 -47 b^4 c+6 b^6 -32
   c^3}{\left(b^2-4 c\right)^2}+\frac{-4 b^2 c+b^4 +2 c^2}{\left(b^2-4 c\right)} +\frac{6 \left(70 b^2 c^2-21 b^4 c+2 b^6-70 c^3\right) w(c, 0)}{\left(4
   c-b^2\right)^2} \right.
 \nonumber \\[4pt]
&&
\left. + 6 \left(2 b^2-c\right) \ln (\delta )+\frac{6 b}{\delta }-\frac{1}{\delta ^2} \right) + O(\delta,z^3).
\label{eqoz0}
\end{eqnarray}

We now proceed to calculate the contribution in the inner region; for this purpose, we use a suitable change of variable as before to perform the integral, obtaining
\begin{eqnarray}
&&I^{int} \equiv \int_0^\delta dx \frac 1{x + bx^2 + cx^3 + z} \stackrel{s \equiv z/x}{=} \int_{z/\delta}^{+\infty} ds \frac s{s^2 + s^3 + b z s + cz^2} %
\nonumber \\[4pt]
&=&
 \int_{z/\delta}^{+\infty} ds
\frac{1}{s+s^2}-z \int_{z/\delta}^{+\infty} ds \frac{b}{s^2(1+s)^2}+ z^2\int_{z/\delta}^{+\infty} ds \frac{(b^2-c-c s)}{s^3(1+s)^3}+ \dots
\nonumber \\[4pt]
&=&
\ln \left(1 + \frac {\delta } z\right)
 -b z \left( \frac{2z\delta + \delta^2}{z(z + \delta)} - 2 \ln
   \left(1 + \frac {\delta }z\right) \right)
\nonumber \\[4pt]
&&-  z^2 \left(\frac{\delta ^2 \left(b^2 \left(\frac{2 z}{\delta }+1\right)
   \left(\frac{6 z \left(\frac{z}{\delta }+1\right)}{\delta
   }-1\right)-\frac{c z \left(\frac{6 z^2}{\delta ^2}+\frac{9 z}{\delta
   }+2\right)}{\delta }+c\right)}{2 z^2 \left(\frac{z}{\delta }+1\right)^2} - 3
   \left(2 b^2-c\right) \ln
   \left(1 + \frac {\delta }z\right) \right)+ \dots
\nonumber \\[4pt]
&=& -\ln \left(\frac{z}{\delta }\right)+z \left(-2 b \ln \left(\frac{z}{\delta
   }\right)-b+\frac{1}{\delta }\right)+z^2 \left(\frac{1}{2} \left(-12 b^2 \ln
   \left(\frac{z}{\delta }\right)-7 b^2 + 2c + 6c \ln \left( \frac z\delta \right) \right)+\frac{3 b}{\delta }-\frac{1}{2 \delta
   ^2}\right)
\nonumber \\[4pt]
&&
   + O(\delta,z^3),
\nonumber \\[4pt]
\label{eqiz0}
\end{eqnarray}
where, as before, we have made a series expansion on $\delta$ up to linear terms for the last equality. The final expression when $z \to 0^+$ comes from adding Eq.~(\ref{eqoz0}) to Eq.~(\ref{eqiz0}). It is not hard to check that contributions involving $\ln \delta$ and inverse powers of $\delta$ cancel out in the final series expansion, and we obtain a result depending only on $z$,
\begin{equation}
f(z)=-\ln z-\ln(b+c+1)-w(c,0)+w(c,1)
+z\left(-2b \ln z-b-\frac{b^3-3 b c }{\left(b^2-4 c\right)} +\frac{2 \left(-6 b^2 c+b^4+6
   c^2\right) w(c, 0) }{b \left(4 c-b^2\right)} \right)
+ \dots
\label{eq:fzsmallFinal}
\end{equation}

\section{Value of $C$ for large-$t$ limit}
\label{appendixb}

In this section, we compute the constant $C$ used in the main text. We start by analysing the behaviour of the integral
\be
F(t) = \int_{f^{-1}(t)}^\infty du\, \left( u - \frac 1{f(u)}\right) f'(u),
\ee
in the large-$t$ limit. For this purpose, we first introduce a cut-off $B$ to obtain
\be
F(t) = \lim_{B\to \infty} \Bigl\{- \ln |f(B)| + \ln t + \int_{f^{-1}(t)}^B du\, u f'(u)  \Bigr\} = \lim_{B\to \infty} \Bigl\{ \ln B + \ln t + \int_{f^{-1}(t)}^B du\,u f'(u) + O(B^{-1})\Bigr\},
\label{bigf2}
\ee
where we used the approximation
\be
f(u) = \int_0^1 \frac 1{x + b x^2 + c x^3 + u} dx = \frac 1u + O(u^{-2}),
\ee
when $u\to \infty$. We evaluate then the integral in Eq.~(\ref{bigf2}) as follows
\begin{eqnarray}
\int_{f^{-1}(t)}^B du\, u f'(u) &=& -\int_0^1 dx \int_{f^{-1}(t)}^B du \frac u{(x + bx^2 + cx^3 + u)^2}
\nonumber \\[4pt]
&=&-\int_0^1 dx \left[ \frac {x + b x^2 + c x^3}{x + b x^2 + c x^3 + u} + \ln(x + b x^2 + c x^3 + u) \right]_{f^{-1}(t)}^B.
\end{eqnarray}
We remark now that in the large-$t$ limit, $f^{-1}(t)$
is of order $e^{-t}$ (see Eq.~(\ref{eq:fzsmallFinal})), and therefore to lowest order, we can set $f^{-1}(t)=0$,
giving,
\be
\int_{f^{-1}(t)}^B du\, u f'(u) = - \int_0^1 dx \left[ \ln B - 1 - \ln (x + b x^2 + c x^3) \right] + O(B^{-1}, e^{-t}) = -\ln B + 1 +\int_0^1 dx \ln (x + b x^2 + c x^3),
\ee
where we drop the $O(B^{-1},e^{-t})$  terms.
The last integral can be found straightforwardly with the result
\be
\int_0^1 dx \ln (x + b x^2 + c x^3) = -2 - \frac {\sqrt{b^2 - 4c}}c \arccoth\left( \frac {b + 2}{\sqrt{b^2 - 4c}} \right) + \frac {(b + 2c)\ln(1 + b + c)}{2c}.
\ee
Putting all together, we arrive at the final expression
\be
F(t) = \ln t - 1 - \frac {\sqrt{b^2 - 4c}}c \arccoth\left( \frac {b + 2}{\sqrt{b^2 - 4c}} \right) + \frac {(b + 2c)\ln(1 + b + c)}{2c}.
\ee
Since we know that $F(t) \sim C + \ln t$ in the large-$t$ limit, we get the expression for $C$ employed in the manuscript.

\bibliography{refartminimal2}

\end{document}